\documentclass{sigchi-arxiv}

\toappear{} 

\usepackage{balance}  
\usepackage{graphics} 
\usepackage{times}    
\usepackage{url}      
\usepackage[percent]{overpic}	

\makeatletter
\def\url@leostyle{%
  \@ifundefined{selectfont}{\def\UrlFont{\sf}}{\def\UrlFont{\small\bf\ttfamily}}}
\makeatother
\def\pprw{8.5in}
\def\pprh{11in}

\usepackage[pdftex]{hyperref}
\hypersetup{
pdftitle={SIGCHI Conference Proceedings Format},
pdfauthor={LaTeX},
pdfkeywords={SIGCHI, proceedings, archival format},
bookmarksnumbered,
pdfstartview={FitH},
colorlinks,
citecolor=black,
filecolor=black,
linkcolor=black,
urlcolor=black,
breaklinks=true,
}

\begin{document}

\title{A Dataset of Naturally Occurring, \\
Whole-Body Background Activity \\
to Reduce Gesture Conflicts}

\numberofauthors{1}
\author{
	\alignauthor Dustin Freeman$^{1}$, Ricardo Jota$^{1}$, Daniel Vogel$^{2}$, Daniel Wigdor$^{1}$, Ravin Balakrishnan$^{1}$  \\
	\affaddr{$^{1}$University of Toronto, Canada}\\
	\affaddr{$^{2}$University of Waterloo, Canada}\\
}

\teaser{
	\includegraphics[width=\textwidth]{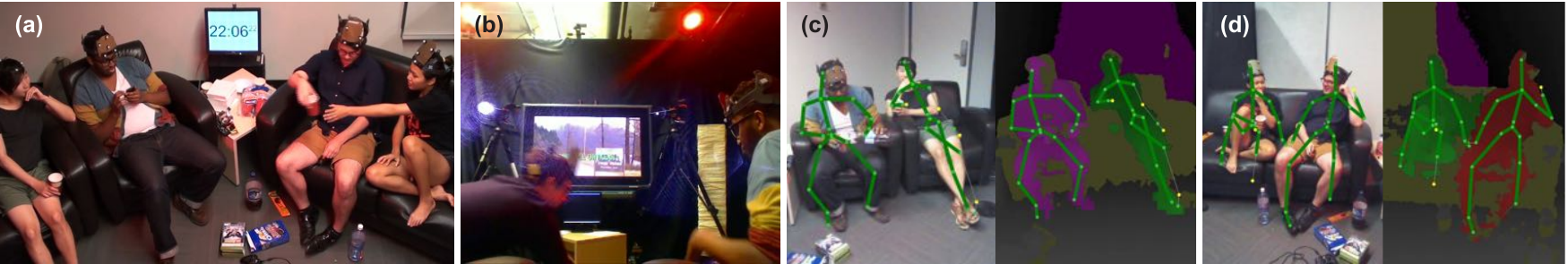}
	\caption{Example living room background activity dataset captured using our tools and methodology: (a) front HD video; (b) rear HD video; (c) Kinect facing chairs; (d) Kinect facing couch. All data is time stamped for synchronization. Kinect steams include colour, depth, skeleton, and spatial audio. Vicon motion capture of head positions (note Òtracking hatsÓ) was included in 7 sessions.}
    	\label{fig:firstpage}
}

\maketitle

\begin{abstract}
In real settings, natural body movements can be erroneously recognized by whole-body input systems as explicit input actions. We call body activity not intended as input actions Òbackground activity.Ó We argue that understanding background activity is crucial to the success of always-available whole-body input in the real world. To operationalize this argument, we contribute a reusable study methodology and software tools to generate standardized background activity datasets composed of data from multiple Kinect cameras, a Vicon tracker, and two high-definition video cameras. Using our methodology, we create an example background activity dataset for a television-oriented living room setting. We use this dataset to demonstrate how it can be used to redesign a gestural interaction vocabulary to minimize conflicts with the real world. The software tools and initial living room dataset are publicly available \footnote{\url{http://www.dgp.toronto.edu/~dustin/backgroundactivity/}}. 
\end{abstract}

\category{H.5.m.}{Information Interfaces and Presentation (e.g. HCI)}{Miscellaneous}

\section{Introduction}

Classifying whether body motions were intended as input is more than just a technological challenge: doing it incorrectly can potentially be deadly. For example, it was recently discovered that people could unintentionally disable their Nest Protect smoke alarm when normal arm movements were erroneously interpreted as a Òwave to silenceÓ gesture \footnote{\url{http://www.bbc.com/news/technology-26879987}}. Misrecognising background activity as an explicit input action is an example of the ÒMidas touchÓ problem \cite{jacko2012human}. Midas touch problems are likely to increase as more always-available whole body input systems are deployed in real environments such as public places \cite{muller2012looking}, classrooms \cite{bolt1980put}, meeting rooms \cite{aggarwal2011human}, and kitchens \cite{panger2012kinect}.

We call naturally occurring activity not intended for input commands \textit{background activity}. Since body tracking and gesture recognition is not yet robust in real environments, the potential for Midas touch problems is compounded. Some types of unexpected or unusual background activity can foul tracking and recognition systems, creating more opportunities for misrecognized input. Avoiding erroneous input is critical to adoption and usability Ð people cannot be expected to carefully constrain their natural motions to avoid misclassification; the problem must be tackled directly. We argue that capturing background activity for observation and design testing is crucial to improving always-available whole-body input. 

In this paper, we contribute a reusable methodology and supporting software tools to generate standardized background activity datasets with 3D motion tracking, depth cameras, spatial audio, and high-definition video (Figure \ref{fig:firstpage}). Our data gathering protocol requires participants to perform explicit prompted gestures at regular intervals, so that datasets contain controlled foreground activity. To validate our methodology, we captured a dataset with 52 person-hours of background activity in a television-oriented living room setting, which we make available to the community.

We ran a proven gesture recognizer for the prompted gestures through our dataset and found a very large number of false positives. This reflects the motivation for our study and dataset: current whole-body interaction design and gesture detection does not consider background activity. As an application of background activity datasets, we design a set of proposed gestures that correspond semantically to our original prompted gestures set. When tested on our dataset, these yield substantially less false positives. We include additional observations about body postures.

\section{Related Work} 

Large datasets of naturally occurring body movements are useful for conducting post hoc observational inquiries, modelling phenomena, motivating technique designs, training algorithms, testing individual techniques, and comparing multiple techniques with a common baseline. Examples of well-established datasets include the MNIST handwritten digit database \cite{lecun1998gradient} for handwriting recognition, the MacKenzie Phrase Set \cite{mackenzie2003phrase} to evaluate text entry techniques, and datasets of static objects captured by depth cameras \cite{janoch2013category,lai2011large} for computer graphics algorithms. Dataset corpora have a strong tradition in natural language processing and have been leveraged to make speech input classification robust to background speech \cite{bohus2009learning,furui2005recognition}. In the field of gesture recognition, algorithms are trained and tested using datasets similar to MarcelÕs \cite{marcel_webpage} compilation of hand gesture and posture images, and to the Cambridge Gesture Database \cite{kim2007tensor} of image sequences showing various hand motions. More recently, the Chalearn gesture challenge dataset was established as part of a competition in ICMI 2012 to recognize gestures consisting of motion and hand shapes in 320x240 Kinect RGB-D data \cite{guyon2012chalearn}.

Datasets of whole-body motion exist, but these focus primarily on short sequences of high-energy actions performed by actors in a motion capture studio \cite{bolt1980put,lei2012fine,ofli2013berkeley,poppe2010survey}. More recently, the CMU Quality of Life Technology Centre created a multimodal capture database of people cooking in a simulated kitchen \cite{cmu_kitchen}. With an average of 5 minutes per clip, the sequences are too short and too task focused to provide general background activity. 

In contrast to pre-existing datasets, we capture much longer sequences with minimally invasive equipment and we encourage a high degree of social interaction and comfort. Rather than clean, segmented sequences of distinct actions, we capture realistic, noisy, everyday actions. Unlike previous datasets, we also intersperse explicit input sequences for baseline testing with natural background activity.

\subsection{Explicit Input with Gestures}
Using body gestures for explicit input has been extensively studied \cite{aggarwal2011human,benko2012miragetable,hilliges2012holodesk,laviola20113d,poppe2010survey,sodhi2012lightguide,turaga2008machine,weinland2011survey}. With always-available body input, the difference between gesture and non-gesture can be subtle, introducing false positives \cite{lei2012fine,panger2012kinect}.  Baudel and Beaudouin-Lafon \cite{Baudel:1993:CRC:159544.159562} call systems that interpret every gesture of the user as possible meaning as having Òimmersion syndromeÓ, ignoring that interacting with the system is not the userÕs only ongoing activity. 

Detecting gestures in a continuous stream of input is known as the Gesture Spotting Problem. A common approach is to model each gesture type as a Hidden Markov Model (HMM) and detect gestures when their likelihood exceeds that of a thresholding HMM, synthesized from the trained gesture HMMs \cite{lee1999hmm}. The limitation of this approach is that this thresholding HMM does not model the background. 

An alternate approach is to design a gesture delimiter that rarely occurs naturally. For pen input, Grossman et al. \cite{grossman2006hover} logged naturally occurring pen hover motions to design distinct hover gestures. For device motion gestures, Ruiz and Li \cite{ruiz2011doubleflip} gathered naturally occurring motion data to design and test the distinct DoubleFlip motion gesture delimiter. These projects demonstrate the use of background activity data, but neither offered a generalizable methodology to capture and distribute the data. To our knowledge, there is no dataset that can be used to evaluate gestural interfaces in the context of naturally occurring whole-body background activity.

\section{Background Activity}

Background activity is interleaved with all interface input, but some input techniques explicitly differentiate between input and non-input actions using an explicit control signal. As a simple example, consider that hand movement is only used for cursor control when a mouse is manipulated --- all other movements away from the mouse are easily ignored. 

When whole-body input systems constantly track the movements of body parts, they can become confounded by the ambiguity between background activity and explicit control. The reason is that control signals can often be very similar to typical background activity movements \cite{panger2012kinect}. An outstretched arm with a pointed index finger could be a gesture to select a location on a computer display (foreground activity) or a deictic gesture to support human communication (background activity). The problem is compounded in active environments where multiple people are multi-tasking with others, or where the physical environment is not conducive to careful, explicit movements.

In computer vision, background subtraction is a common method to separate objects of interest using a model of the image background \cite{stauffer1999adaptive}. The separation of foreground objects (explicit input) is achieved by a deep understanding of the background scene (i.e., background activity). We argue that the whole-body input research can use an analogous approach. Current gesture and motion training datasets \cite{cmu_mocap,shotton2013real} are not suitable; a corpus of background activity datasets in realistic environments is needed, as are a methodology and tools to enable collection of additional datasets.

\subsection{Approaches to Managing Background Activity}

There are multiple gesture detection approaches to distinguish foreground activity from noisy background activity. We present the most common approaches:

\emph{Explicit Clutch} --- The system only responds when in a specific user-determined state. For example, Sapponas et al. \cite{saponas2009enabling} use a clenched left fist to enter a gesture recognition state for the right hand.

\emph{Delimiter} --- A special gesture indicates that an input sequence is about to begin. Sometimes the delimiter is multi-modal \cite{bolt1980put}, but when using pure whole-body input, a unique gesture can be used \cite{walter2013strikeapose}. There are varying ways to indicate the end of the interaction sequence, such as ending after the first gesture recognized \cite{ruiz2011doubleflip}, or after a period of inactivity. Hudson et. al. \cite{Hudson:2010:WGI:1709886.1709906} use the term framing gesture when a delimiter is performed before and after the interaction sequence.

\emph{Implicit Clutch} --- The system examines the gesturing context to determine when a hand motion should be interpreted as input. Baudel and Beaudouin-Lafon \cite{Baudel:1993:CRC:159544.159562} and Fourney \cite{fourney_msc} use a spatial active zone: when the hand is in a certain area, the movements are recognized.  Other features can be used, such as body pose and gaze \cite{Schwarz:2014:CBP:2556288.2556989}, to determine when hand gestures are intended as input.  

\emph{Always-On} --- The system constantly tracks and responds to any motions that are intended for input. This requires motions to be clearly distinguished from background activity. The challenge is to find unique, distinct gestures. This is the approach taken by the Nest Protect smoke alarm.

There are benefits and issues with these approaches. An explicit clutch is clear, but requires user attention to be maintained. Delimiters do not have to be maintained during the interaction, but require extra time at the beginning. Assuming a robust clutch or delimiter can be found, it can still feel awkward and require extra cognitive effort to use. Another problem presented by a strict delimiter approach is that it precludes detecting movements that could be used as implicit input, like sensing emotional affect \cite{aggarwal2011human} or level of attention \cite{vogel2004interactive}.

Always-on and implicit clutch are the most natural modes of interaction. To be usable and reliable, they require the deepest understanding of background activity. The active zone implicit clutch works well in a presentation context, but is unlikely to generalize. The hand wave used by the always-on Nest Protect was not unique enough. 

Our approach is to study background activity to move closer to the goals of implicit clutches and always-on input. In the next section we describe a principled way to capture background activity datasets. We then demonstrate the usefulness of our initial dataset to find unique and robust gestures that do not frequently occur in background activity. This approach could be combined with others, for example these unique gestures could be used as a delimiter as well and the dataset could be used to design and test implicit clutches. 

\subsection{Establishing a Methodology for Dataset Building}

Our objective is to establish a repeatable methodology for capturing an ecologically valid recording of whole-body background activity in a form suitable for distribution. In this section we establish a study protocol that includes occasional prompted foreground activity segments for baseline comparison, provide format specifications for a public domain dataset, and describe our logging and analysis software. We use our methodology to capture background activity in a television-oriented living room, a plausible context for whole-body interaction. 

\begin{figure}[th]
  \begin{center}
  \includegraphics[width=0.9\columnwidth]{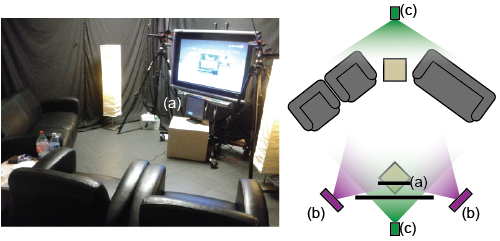}
  \caption{Living room environment with seating and large screen television. (a) small display for prompted foreground activity gestures; (b) Kinect cameras; (c) HD cameras.}
  \label{fig:living_room}
  \end{center}
\end{figure}

\subsection{Eliciting Background Activity}

Unlike typical methodologies where people are instructed to perform specific motions, asking people to Òact outÓ background activity would not produce realistic results. We therefore advocate creating a physical and social environment that allows background activity to emerge naturally. For our sample dataset, we created a laboratory living room setting with a game console and television (Figure \ref{fig:living_room}). To increase social interaction, we only recruited participants who had existing social relationships with each other. To encourage object manipulation background activity, we provided snacks and drinks.

To gain full benefit from the dataset, the inclusion of typical foreground gesture activity is essential to serve as a comparative baseline. We achieve this by occasionally prompting participants in a subset of groups to perform one of four common gestures. Our methodology may be extended for testing a particular gesture language, by adding those gestures to the experimental protocol. This can even be done before recognizers have been built for those gestures and used to inform the recognizer implementation.

\section{Capture Protocol}
\subsection{Physical Environment Setup}

We simulated a 4 m by 4 m living room with comfortable furniture and used soft incandescent lighting and curtains to hide the institutional walls (Figure \ref{fig:living_room}). We placed two armchairs and a two-person sofa in front of a 54'' television with external speakers approximately 2m away. Participants could watch Netflix programs or play video games, controlled using a single wired Xbox controller. We intentionally provided a single controller to increase background activity: controller usage had to be socially negotiated and transferred. Similarly, background activity was encouraged when selecting a video game from a stack on the floor and inserting it into the Xbox. 

To maintain an unobstructed view of the participants, we placed a small coffee table between the couch and the nearest armchair, rather than in front. This table held food and other personal belongings within armÕs reach of the two nearest participants. This was another intentional choice to encourage background activity, since items on the table needed to be passed to the two outer participants.

\vfill
\subsection{Capturing Apparatus}

We used minimally invasive capture equipment to capture each study group of 4 participants. A wide-angle HD video camera captured audio and video of the entire scene from the front (Figure \ref{fig:firstpage}a) and a second HD camera captured from behind, including the gesture prompt screen and television content (Figure \ref{fig:firstpage}b). 

One Kinect faced the sofa (Figure \ref{fig:firstpage}d), and the other faced the armchairs (Figure \ref{fig:firstpage}c). Each Kinect recorded 13 bit, 640 by 480px depth with 3 bits of player id masks (pixels classified as part of a human body), 640 by 480px RGB video, 20 segment skeleton tracking (when possible), and spatially separated sound using Microsoft Kinect SDK version 1.5. When used, a six-camera Vicon system placed high in the ceiling tracked head position and orientation of all four participants using four lightweight hats. We were concerned that the Vicon Òtracking hatsÓ would affect behaviour, so we used them with only a subset of groups in order to increase the breadth of our sample dataset. 

We found that the built-in Kinect SDK recorder produced extremely large files (typically 1.5 GB/min per Kinect). To keep the data manageable, we designed a more efficient capture format (typically 0.3 GB/min). We used RIFF as a generic container to house all time-indexed depth, RGB, and skeleton frames in one file. RGB frames were compressed with lossy JPEG compression and depth frames with lossless LZF compression. Since the Kinect SDK does not output depth, RGB, and skeletal frames at a consistent rate, each frame is time stamped. We provide Windows C\# software to capture and playback Kinect data in this format, as well as Python software for gesture detection and other analyses. We plan to update the file format and tools for the Microsoft Kinect 2. A detailed file format description is included with the dataset to enable other implementations. 

\subsection{Public Dataset Concerns}

We were careful to gain approval from our research ethics board so that we could make the dataset publicly available. Participants were warned of this in advance of arriving at the study, and were given 1 week after the study to contact the researcher if they had concerned. To ensure the dataset is rich and useful in analyzing background activity, full audio is included, and faces will not be blurred. While the details of the dataset are be publicly available, its full download is only be possible after a Terms of Use is agreed to, identifying the dataset user as a researcher.

\subsection{Participants}

A large amount of background activity is socially motivated, so we recruited participants in groups instead of individuals. Online posting and word-of-mouth yielded 13 groups of four participants, for a total of 52 participants. The mean age was 26 years (ranging from 19 to 59). Overall, 67\% of our participants were male, but gender distribution within groups varied: one all-female, four all-male, and the remaining mixed. Seven groups used Vicon motion tracking, seven groups included prompted foreground gestures, and five groups had both. 

In three groups, one participant was meeting the others for the first time, but all others had existing social relationships. Pairs of participants with closer relationships would often rush to the sofa and were often physically affectionate. In one group, one participant was frustrated with the other members and avoided social interaction Ð he spent most of his time reading a newspaper. 

\subsection{Procedure}
The procedure emphasizes putting participants into a mood suitable for the simulated environment. In the case of our living room simulation, this meant getting participants comfortable and minimizing the feeling of being in a lab. The researcher always met participants outside the building and guided them to the study room on a route planned to minimize time in office spaces. During the walk, the group was engaged in small talk to help everyone relax. We wanted participants to act as if at home Ð shouting, cheering, joking Ð without worrying about disturbing others working in the building. Study times also reflected this social situation, with most group captures occurring in the evenings or on weekends. 

To increase background activity, food and drinks were placed on the coffee table in the study environment, along with disposable plates, cups, and napkins, and a garbage can. Participants were told to help themselves to the snacks. 

In instances where prompted gestures were collected, the researcher gave instructions on performing them (details below). He then provided instructions on the use of the Xbox console. Participants were encouraged to relax and enjoy whatever they wished on the television, or to just talk, as long as they remained in the simulated living room space and in the same order on the furniture.  The study ran for 60 minutes. During this time, the researcher remained out-of-sight in a nearby location monitoring the capture streams in case there were any problems, and then gave a five-minute warning before the study ended.

\subsubsection{Prompted Foreground Gestures}
To capture the difference between background activity and intentional gestures, we selected a set of gestures to be prompted during the session. We chose four common gestures: Horizontal Swipe, whole-hand AirTap \cite{vogel2005distant}, Wave \cite{vogel2004interactive}, and Point \cite{fourney_msc} (Figure \ref{fig:gestures}). Horizontal swipe is a left or right motion ($\sim$60cm) with the palm perpendicular to the large display, arm extended away from the body, and elbow relaxed. AirTap is a forward and back movement ($\sim$25cm) with palm facing the large display. Wave is a left and right periodic motion ($\sim$25cm) with the elbow roughly fixed in space. Point extends the arm and index finger towards the television. The required duration of Wave and Point were 800ms. These gestures were chosen since they have been used for explicit input, with demonstrated successful detection, but we believed they were also likely to occur in background activity. We kept the set of gestures small to reduce cognitive load on our participants and avoid interference with our primary goal of observing background activity. 

\clearpage
Seven groups (out of 13) were regularly prompted to perform gestures to capture foreground activity in the context of background activity. A 17-inch display below the television (Figure 2a) prompted people to perform a one of the prompted gestures using an iconic representation and audio cue. We prompted participants by number (1-4), and each performed each gesture at least once. The prompt displayed until the gesture was ÒrecognizedÓ by the researcher monitoring the camera feed (i.e., a Wizard-of-Oz recognition technique). 

The only feedback provided is that the prompt will disappear when the gesture is correctly detected. It would be difficult to provide high-fidelity feedback when our elicitation procedure is Wizard-of-Oz, and it would have prevented us from accepting complex interleavings of foreground and background activity. While feedback could possibly improve our true positive detection rate, the primary goal of this study is to collect background activity, which would not be affected by any sort of feedback.

Before the study began, the researcher demonstrated each gesture to the group twice. The researcher left the room so that each participant could practice following the small display prompt to perform one gesture. All gesture-training demonstrations are included in the dataset. Each gesture was prompted five times during the 60-minute session, resulting in a foreground gesture sequence approximately every three minutes.

\section{Results}

We captured 1 hour of data per group of 4 people, totalling 52 person-hours of background activity and 750 GB of data.

\subsection{Participant Behaviour}

Most groups played a game or watched television while also talking, eating, and using mobile devices. While the television display was the primary focus, participants were almost always multi-tasking. Participants assumed a wide variety of comfortable positions on the furniture that suggest we were successful at simulating a realistic living room setting.

Intensity of background activity varied. Aggressive gesticulation was common, especially for boisterous groups. One group of hip-hop dancers was very expressive with a high level of dynamic movement. Another group had two of its members subconsciously compete to be the center of attention, outdoing each other in speaking volume and gesticulation intensity. There were also quieter groups, such as a married couple and one set of parents. This group quietly watched a movie and ate snacks, speaking occasionally.

\subsection{Prompted Gestures}

For groups with prompted gestures, we captured a total of 140 gesture sequences (7 groups x 4 gestures x 5 prompts). We noticed that well-intending participants reminded others to perform a gesture. This usually involved some communicative gesture similar to the required gesture. Nonetheless, because this appeared to be an artefact of our setting, we asked participants not to engage in this behaviour. 

\vfill
\subsection{Capture Quality}

The Kinect captured data at between 15-30 fps. For groups with Vicon motion tracking, 6 DOF data for each hat was captured at between 60-120 fps. At first the tracking hats seemed conspicuous to some of the participants, but they relaxed after 10 minutes or so. Hat tracking data is included in the full data set, despite not being included in this paper.

\section{Recognition of Prompted Gestures}

Background activity datasets can be used to test different gestures and recognizers. As an example, we use our initial dataset to evaluate the performance of a HMM Gesture Spotting Network (GSN) with the four prompted gestures: Swipe, Point, Wave, and AirTap. These results are dependent on skeletal tracking quality for hand position, a realistic limitation when using current skeletal trackers, especially in environments like a living room, where relaxed postures might cause poor skeleton tracking. 

\subsection{HMM GSN Design, Implementation, and Training}

Our design is based on Fourney \cite{fourney_msc} and Lee and Kim \cite{lee1999hmm}. A GSN is a meta-Hidden-Markov-Model (HMM) containing multiple HMMs connected in parallel. There are left-to-right gesture HMMs for each variation of the gesture to be detected and a special threshold HMM representing non-gesture movements. A gesture is detected (or ÒspottedÓ) when the final state of one of the gesture HMMs has a higher likelihood than every state in the threshold HMM. Our left-to-right gesture HMMs were constructed of 4 states each.

Like Fourney, we discretize body-relative hand position and velocity into features, although ours are in 3D. We designed features by plotting the training gestures and determining how best to distinguish between them. We measure the depth of the hand relative to the shoulder and its horizontal and vertical position relative to the elbow. We take the 3D position and assign four discrete features: one binary thresholded radius, as well as three features for angles between the hand vector and depth sensor-relative axes. The angle features each have three possible values, of the form $(\theta < -\pi/4, - \pi/4 < \theta < \pi/4, or \pi/4 < \theta)$.  We found spherical coordinates were a good model for the 3D hand position relative to the body, as suggested by Freeman et. al \cite{Freeman:2014:LLA:2611105.2557304}. The discretized velocity feature is the nearest 3D unit vector of form [\{-1,0,+1\},\{-1,0,+1\},\{-1,0,+1\}].

We found that participants performed Swipe and AirTap with a few variations, so a single HMM could not  describe these gestures. Instead, we trained one gesture HMM for each variant. Swipe has two variants: elbow straight and elbow bent. AirTap motions all began with a quick forward motion, but finish with one of three variants: relaxing the arm, dropping the arm, or pulling the arm back quickly. 

\begin{figure*}[!t]
  \begin{center}
  \includegraphics[width=\textwidth]{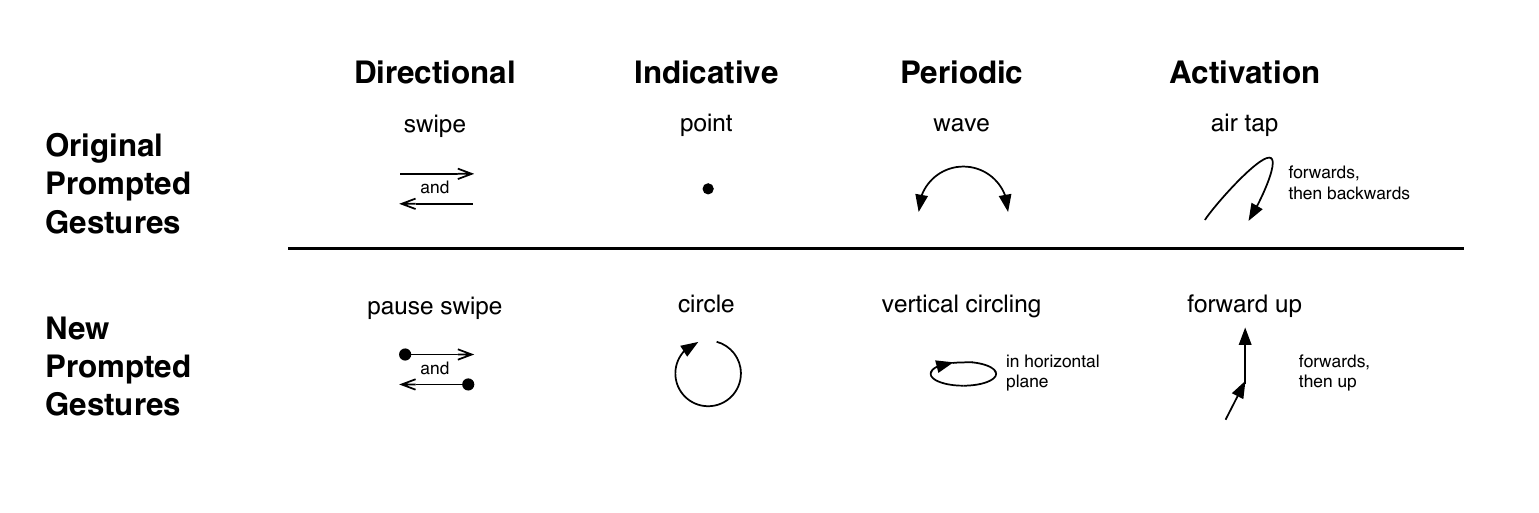}
  \caption{Diagrammatic representations of the original prompted gestures used in the capture study, followed by the corresponding proposed gestures developed using the dataset. They are semantically similar but with substantially less false positives.}
  \label{fig:gestures}
  \end{center}
\end{figure*}

For training data, six volunteers, who did not participate in the study, performed each gesture variant 3 to 15 times while seated. After discarding approximately 20\% of cases with poor tracking or unusual motions, there were 80 training examples per gesture variant. We trained the gesture HMMs using the Baum-Welch algorithm, with 10\% of the training examples as held-out test data. 

We found 99\% accuracy on the test data by comparing likelihood of individual gesture HMMs. When we constructed a GSN by adding a thresholding background model, accuracy reduced to 83\%. The worst performing gesture was Point (79\%), which was often incorrectly recognized as background (13\%). As the Point gesture is simply the user holding still in the same state, this is difficult to distinguish from the same behaviour in the background model by the standard method of background model construction.

\subsection{Results}

We first evaluate the true-positive rate using the 140 prompted gesture sequences. Given noisy data, with participants sitting in relaxed postures often quite close to each other and blocked by props such as food items or controllers, the performance of the Microsoft Kinect SDK (v1.5) Skeleton tracker was affected. We manually examined the skeletal data on each prompted gesture sequence and found only 44\% of these to have decent skeletal tracking. For these, the true positive detection rate for prompted gestures was 48\%. 

While the detection rate appears low, we consider this fairly good, given the diversity of gesture performance, examples of which are in the video accompanying this paper. The participants did not train against a recognizer, and the researcher was intentionally liberal with their definition of a correct gesture in the Wizard-of-Oz procedure. Over the duration of the study, gesture performance became subtler, more individualized. We frequently saw participants repeating a gesture multiple times in quick succession. Swipe and AirTap in particular changed substantially over time. We think it is important to not just study more realistic gestural background activity, but more realistic performance of gestures when participants are tired, or even bored. Indeed, Negulescu et. al have proposed using a second, lower threshold for recognition when two barely-recognizable gestures are performed immediately after one another \cite{negulescu2012recognition}.

To evaluate false-positive rates, we ran the GSN over each tracked skeleton in all background activity sequences. We found 73,729 false positives: 38,005 for Swipe, 15,716 for Wave, 19,120 for Point, and 888 for AirTap. In total, this is one false positive every 5.1 seconds per-participant. We examined 20 false positives for each gesture and found many cases where poor skeleton tracking was the cause. The results indicate that our proposed gestures are abundant in background activity, which results in a high false positive recognition rate even with a reasonable true-positive detection rate. 

Focusing on false-positives with good skeletal tracking, we identified five common causes: reaching or manipulating objects, gesticulating, touching, repositioning, and stretching. Reaching or manipulating an object created motions similar to a point or swipe. Gesticulation led to expressive hand movements that could look like any of the gestures. When participants touched themselves, such as scratching, a wave gesture was often recognized. When participants repositioned their body, such as leaning back and extending their arms forward on the armrest, this appeared as a forward-extended point gesture. Finally, stretching, often with both arms, triggered an AirTap or forward point gesture. In the next section we discuss design implications based on these causes to reduce these false positives. This is only an initial examination of false-positive causes; the dataset provides the means to complete a more formal analysis.

As we note before, none of these actions are avoidable in the real world. Regardless of how successful the recognizer is in identifying these gestures, they will always be susceptible to misrecognitions. What we need are gestures that are still reasonable to perform but also unique, in the sense that they do not frequently appear in the background activity.

\section{Proposing New Gestures}

The prompted gestures we na{\"i}vely chose produced far too many false positives to be useful in a real scenario. While recognition may be improved by continually researching a better recognizer, this will provide diminishing returns. We demonstrate the utility of background activity datasets by using our living room dataset to redesign our gesture set to be more robust to the real-world activity, without any changes to the design of our gesture recognizer.

To test the utility of a given gesture in a certain background activity context, we can simply train a detector to recognize the gesture, then run it through our data and count the number of false positives, where fewer false positives is better. This is an extension of previous procedures used in different sensing domains \cite{ruiz2011doubleflip}.

We created a set of proposed gestures that semantically correspond to each gesture in our prompted gesture set (Figure \ref{fig:gestures}). Instead of left and right swipe, we create Pause Swipe, a swipe that is preceded by a short pause; this preserves the swipesÕ directional property. Instead of point, we create Circle, meant to be a single circle motion of the extended arm parallel to the torso of at least 30 cm in radius; this preserves the point gesturesÕ ability to indicate an object by circling around it, as if with a cursor. Instead of wave, we create Vertical Circling a continuous circling motion in the horizontal plane with the arm extended upwards from the elbow; this preserves the periodic property of wave, providing a gesture that could be performed until a system response is given. Instead of AirTap, we implement Forward Up, a push forward towards the interface, then an upward flick. This preserves AirTap's diectic sense that a specific location on the surface is being activated or approved, similar to a click.

We trained our gesture recognizer on 10 examples of each of these proposed gestures. We ran our same GSN HMM recognizer through the dataset to look for these gestures, and consistently found fewer false positives. For Pause Swipe, we found 2,494 false positives (15.2 times less than Swipe); for Circle, we found 5,409 false positives (3.5 times less than Point); for Vertical Circling, we found 5,172 false positives (3 times less than Wave); and for Forward Up, we found 268 false positives (3.3 times less than AirTap). Overall, we reduced the false positive rate by a factor of 5.5.

We have successfully produced gestures that are not difficult to perform, yet are far less common in background activity. While we have only created a tested a single alternative to each original gesture here, this methodology could be fused with other approaches, such as implicit clutching.

\section{Qualitative Observations}

\subsection{Body Postures}

A corpus of background data can be used to classify natural postures in a given setting. Here, our goal is to classify body postures that occur in a comfortable environment like the living room. These can be individual postures or combined to include multiple bodies. Our results are relevant to understanding the availability of a person's specific body parts to provide explicit input for a computer system, which could aid in off-line gesture design, as well which type of controls the system offers in-the-moment. It is also possible that this could motivate a model of typical movements, given a certain body posture - this would allow a system to better distinguish unusual movement (a candidate for foreground activity) from background activity. In addition, this provides motivation for improving body and skeletal tracking for this kind of environment.

To find static postures, we used a script to extract depth and RGB frames from the data where the depth frames had inter-frame differences below a threshold for five seconds or more. This resulted in 2014 frames from the two scenes (couch and chairs). The frame samples are reasonably uniform across studies, with a median of 51 samples for the two scenes across 13 groups. Using these frames, we classified postures according to two characteristics: torso lean and arm position. We also observed interesting multi-person body postures.

\begin{figure}[!h]
  \includegraphics[width=\columnwidth]{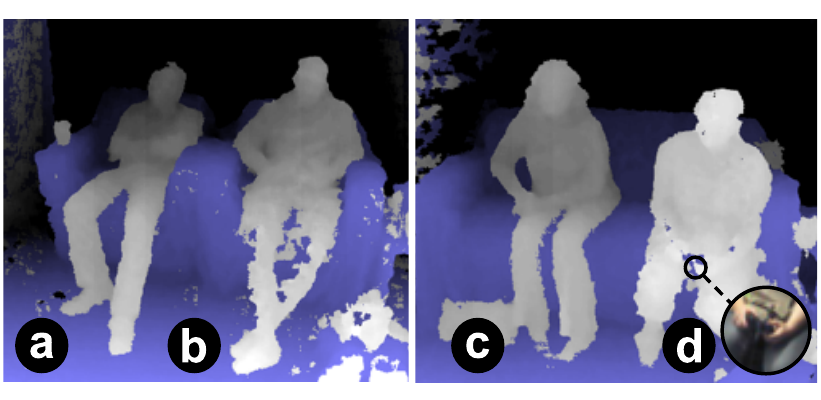}
  \caption{Torso lean degrees: (a,b) backward lean (least active); (c) neutral lean; (d) forward lean (most active).}
  \label{fig:torso_lean}
\end{figure}

\subsubsection{Torso Lean}

We found that the degree of torso lean is a useful way to gauge how available someone is for performing explicit input. We categorized leans into three levels. In decreasing level of availability: forward, neutral, and back (Figure~\ref{fig:torso_lean}).

A \emph{forward lean} is when the head and shoulders are in front of the hips; arms have less contact with furniture, and attention focus is forward. This often resulted from handling food, mobile devices, or the Xbox controller. 

A \emph{neutral lean} when the torso is near vertical; arms on armrests with one arm often supporting the head. In this case, one arm typically remains available for interaction. 

A \emph{backward lean} is characterized by the body appearing relaxed, with the torso fully supported by the backrest, often adopting asymmetrical poses with crossed arms and legs. This is the least probable torso lean for interaction.

\begin{figure*}[!t]
  \begin{center}
  \includegraphics[width=0.7\textwidth]{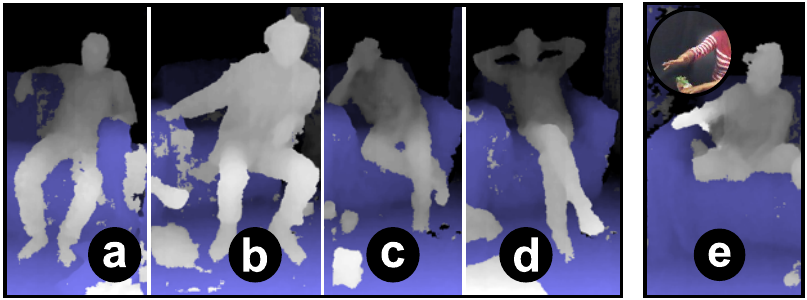}
  \caption{(a) - (d) Examples of arm unavailability: (b) Participant gesturing with the available hand. Note, in the RGB overlay, the other hand is occupied with a bag of chips.}
  \label{fig:arm_unavailability}
  \end{center}
\end{figure*}

\begin{figure*}[!t]
  \begin{center}
  \includegraphics[width=0.6\textwidth]{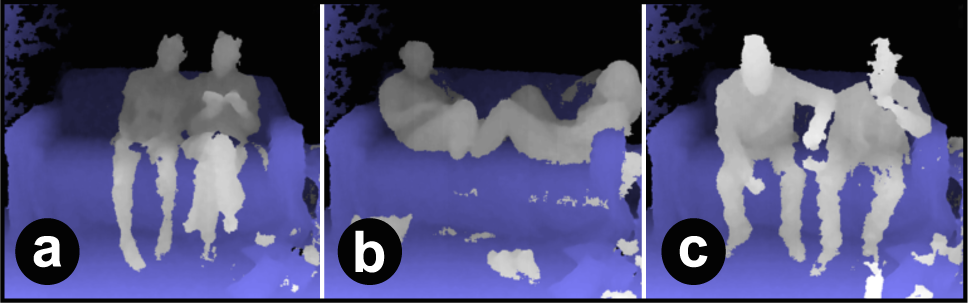}
  \caption{Examples of combined body postures: (a) pressing torsos together; (b) interweaving legs; (c) sharing food.}
  \label{fig:combined_body}
  \end{center}
\end{figure*}

\subsubsection{Arms}

We observed a variety of different arm postures, ranging from extended arms far away from the torso, to crossed arms, and arms kept close to the body. Body symmetry is indicative of which limbs are available for performing explicit input motions. Any limb supporting the body, head, or other objects is unavailable for immediate explicit input. Even when resting, relaxed extended arms, aimed towards the system were indicative of availability (Figure~\ref{fig:arm_unavailability}).

\subsubsection{Combined Body Postures}

We observed combined body postures where two people sat  close. This happened when sharing food, viewing another person's mobile device and expressing intimacy. In these cases, skeletal and gesture recognizers' effectiveness was very low. Gesture designers could specifically consider close postures, for example, designing two-person gestures (Figure~\ref{fig:combined_body}).

\subsection{Qualitative Evaluation: Body and Skeleton Tracking}

We used our dataset to evaluate Kinect SDK tracking. We found that the tracker performs well when people sit upright and make large movements, but performs poorly when people are seated with legs crossed, leaning, touching other people, or holding objects. To investigate methodically, we reviewed the 140 prompted gesture sequences. 

We found 62 (44\%) of these sequences have properly tracked skeletons. Due to issues with low or uneven depth frame rates or lack of skeleton recognizer output, 41 sequences (29\%) have no skeletal data. However, the depth data quality in 33 of these sequences should be adequate for post-capture skeletal detection using other libraries.

The remaining 37 (26\%) of the sequences represent interesting failure cases. In five sequences (4\%), the participant was sitting in a position that makes skeleton detection difficult, such as having their legs crossed or arms folded tightly (see body posture observations above). In 15 cases (11\%), the skeleton was generally correct, but another object was erroneously tracked as the dominant hand (often the participant's torso, leg, or parts of the furniture). This failure was likely due to the arms being held close to the body or hands occupying a small area when extended directly towards the camera. In 11 cases (8\%), a skeleton was detected away from the two primary participants in the scene, such as on some of the items in front of the participants, or another participant leaning into frame. Since the Microsoft Kinect SDK supports a maximum of two skeletons simultaneously the addition of this new skeleton resulted in an inability to track the participant performing the prompted gesture. For six cases (4\%), person-tracking merged two people sitting close together, creating aberrant skeletons. This was most pronounced in one session where a couple sat close together on the couch. Two of the sessions without prompted gestures also have sequences where body tracking merges people sitting close together. Identifying and correcting these failure cases has the potential to improve tracking.

\begin{figure}[h!]
  \centering
  \includegraphics[width=\columnwidth]{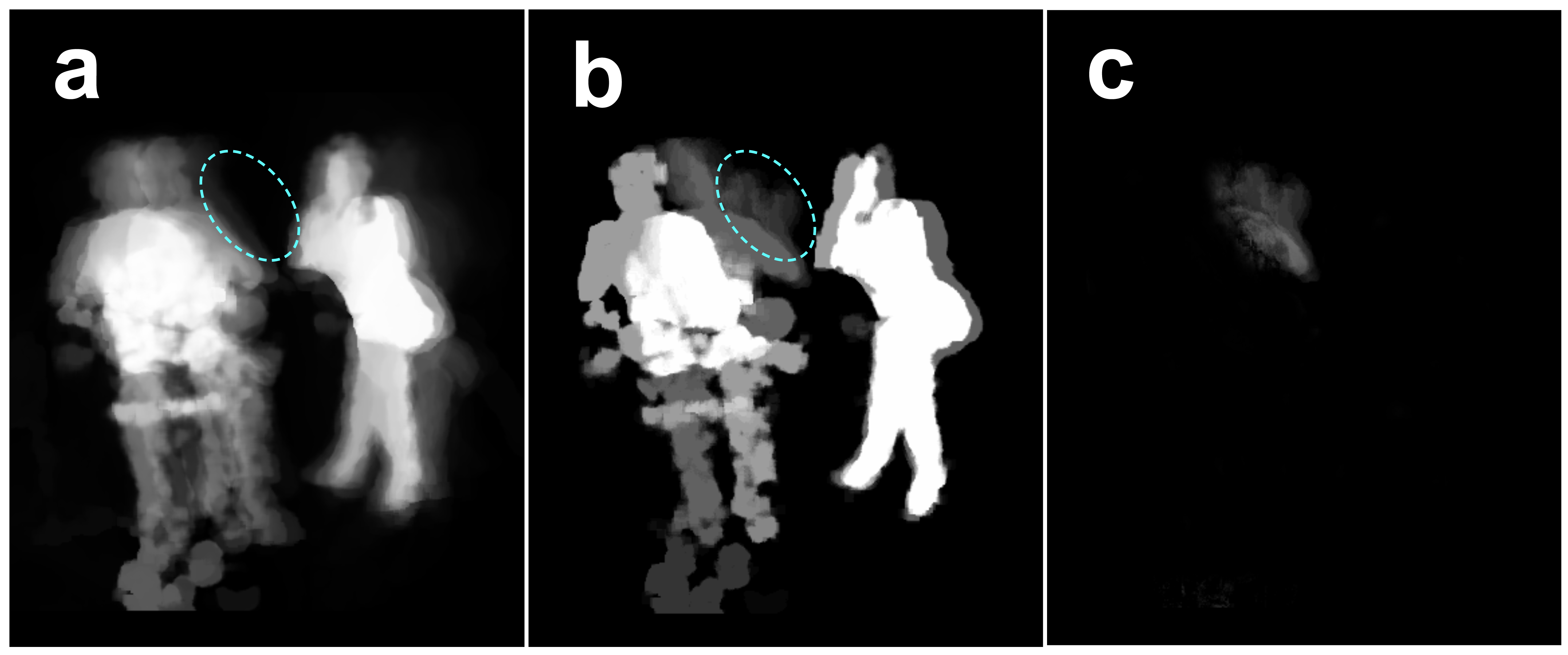}
  \caption{Proposed gesture-specific spatial zones visualized using average depth occupancy: (a) background sequences; (b) AirTap gesture sequences; (c) subtraction revealing spatial gesture zone. }
  \label{fig:spatial_zones}
\end{figure}

\subsection{Gesture-specific spatial zones}

We observed participants performing gestures at greater distances from their body than typical background motions. To operationalize this, we calculated the average body depth during background sequences (Figure~\ref{fig:spatial_zones}a) and average body depth occupancy during prompted gesture sequences for each type of gesture (for AirTap, Figure~\ref{fig:spatial_zones}b). Subtracting the average background occupancy from average gesture occupancy reveals a spatial zone where that gesture was performed. Although they appear similar, early results indicate that gestures may populate spaces not common to background activity. 

\vfill
\section{Conclusion and Future Work}

We described a methodology to capture whole-body background activity and use it to capture a television-oriented living room dataset. To demonstrate the utility of this approach, we use the dataset to redesign a gesture set, and substantially reduce false positives found by a Hidden Markov Model-based Gesture Spotting Network recognizer. A major novelty of this dataset is that it interleaves controlled, prompted foreground activity with long periods of multi-person, open-ended background activity Ð making this kind of analysis possible. Our documentation of this process includes critical aspects that would be necessary in future work, including social considerations, ways to prime activity, and the effect of furniture placement.

These practical findings are encouraging, but it is important to note that our living room dataset and example dataset applications are primarily intended to illustrate and validate our reusable capture methodology. In particular, the large amount of rich data recorded, containing a variety of realistic tasks, could be used to further explore implicit clutching, natural poses, social interaction, etc. The living room dataset and supporting capture and analysis tools are made available to the research community. 

Our primary contribution is to call attention to background activity, which has been under-studied and under-acknowledged in whole-body gestural interfaces appearing in the research community. While it is often not feasible to explore background activity at the very early stages of interaction technique development, it is an important second step to fully understand this new interaction paradigm. It would be ideal if there was a context-independent set of motions characterizing all background activity. While there may be some commonalities, our data collection was only in a living room context. This is arguably a critical context to study background activity given many home entertainment applications, but making any claim of generalising background activity across contexts is premature.

Our intention is that these methods, tools, and techniques presented will assist in the research and design of whole body gestural interactive systems by motivating the capture and sharing of many background activity datasets. In addition, our work provides encouraging results for the design of new always-on gestures. This supports our argument that understanding background activity is crucial to bringing always-available whole-body input into the real world.

\section{Acknowledgements}

We acknowledge members of the Dynamic Graphics Project for various assistance, especially John Hancock. We also acknowledge our study participants for taking part in this novel study.

\bibliographystyle{acm-sigchi}
\bibliography{ba_arxiv}
\end{document}